**FRONT MATTER**

## Title

# Modulation-Frequency Dependence of Spatial Resolution in Optical Correlation-Domain Reflectometry

## Authors

Keisuke Motoda[1], Takaki Kiyozumi[1,2], and Yosuke Mizuno[1,3*]

## Affiliations

[1] Faculty of Engineering, Yokohama National University, Yokohama, Japan.

[2] Graduate School of Engineering, The University of Tokyo, Tokyo, Japan.

[3] Institute for Multidisciplinary Sciences, Yokohama National University, Yokohama, Japan.

Address correspondence to: Yosuke Mizuno; mizuno-yosuke-rg@ynu.ac.jp

## Abstract

The spatial resolution of optical correlation-domain reflectometry (OCDR) has conventionally been described by an expression that is independent of the modulation frequency $f_{\mathrm{m}}$, source linewidth $\delta\nu$, and receiver resolution bandwidth (RBW). However, our previous measurements showed that the spatial resolution $\Delta z$ improves with increasing $f_{\mathrm{m}}$. Here, we develop a theoretical model for OCDR with a frequency shifter by evaluating the electrical power detected by an electrical spectrum analyzer and explicitly including $\delta\nu$ and the RBW $B$. The model predicts two regimes. At low $f_{\mathrm{m}}$, $\Delta z$ decreases approximately in proportion to $1/f_{\mathrm{m}}$, with the proportionality determined by the combined source and receiver spectral response. At high $f_{\mathrm{m}}$, $\Delta z$ approaches a constant value determined by the modulation amplitude $\Delta f$ and independent of the source linewidth and RBW filter. Measurements at RBW = 1 MHz reproduced the transition between these regimes over correlation orders up to 2048. At RBW = 10 MHz, a Voigt representation of the spectral response overestimated $\Delta z$, whereas direct use of the measured unmodulated beat spectrum reduced the discrepancy to approximately 10 to 20%. These results provide a quantitative description of the modulation-frequency dependence of OCDR spatial resolution and clarify its trade-off with measurement range.

## INTRODUCTION

The continued expansion of high-capacity optical communication networks and distributed fiber-optic sensing systems has increased the demand for reliable techniques capable of spatially resolving reflection, loss, strain, and temperature along optical fibers. Representative fiber-reflectometry techniques include optical time-domain reflectometry (OTDR) [1,2,3], optical frequency-domain reflectometry (OFDR) [4,5,6], and optical correlation-domain reflectometry (OCDR) [7,8,9,10,11]. The same correlation-domain principle has also been extended to distributed Brillouin sensing through Brillouin optical correlation-domain reflectometry and analysis (BOCDR/BOCDA) [12,13,14,15,16,17]. OTDR enables measurements over long distances but generally provides relatively coarse spatial resolution, whereas OFDR offers substantially higher spatial resolution at the expense of measurement range and requires a narrow-linewidth light source and a highly stable interferometric configuration. OCDR, based on the synthesis of optical coherence function (SOCF) [8,9], occupies an intermediate regime, combining comparatively long measurement range with moderate spatial resolution, real-time operation, and random accessibility to arbitrary positions along a fiber from a single end. These characteristics make OCDR particularly attractive for distributed diagnostics and sensing applications in which measurement range, spatial resolution, and system complexity must be balanced.

In OCDR, sinusoidal frequency modulation of the light source produces periodic correlation peaks along the fiber under test (FUT), and the measurement range is inversely proportional to the modulation frequency $f_{\mathrm{m}}$ [8,9,18]. The spatial resolution $\Delta z$ has conventionally been defined by the full width at half maximum (FWHM) of the synthesized optical coherence function as $\Delta z = 0.76c/(\pi n \Delta f)$, where $c$ is the speed of light in vacuum, $n$ is the effective refractive index of the fiber, and $\Delta f$ is the modulation amplitude [9,18]. This expression contains neither $f_{\mathrm{m}}$, the source linewidth $\delta\nu$, nor the resolution bandwidth (RBW) of the electrical spectrum analyzer (ESA). Thus, the conventional theory predicts that $\Delta z$ is independent of $f_{\mathrm{m}}$ and the receiver bandwidth. Previous work has examined the spatial resolution of simplified OCDR (S-OCDR), which does not use a frequency shifter [18]. Other studies have considered arbitrary modulation waveforms in BOCDR [19] and strain and bandwidth effects in Brillouin correlation-domain sensing [20,21]. In S-OCDR, zero-span detection must be offset from the beat center to avoid low-frequency noise. As $f_{\mathrm{m}}$ changes, the modulation sideband falling within the detection band also changes, resulting in an $f_{\mathrm{m}}$-dependent spatial resolution [18]. This mechanism does not apply to OCDR with a frequency shifter, where an acousto-optic modulator (AOM) provides a fixed frequency shift and the zeroth-order component is detected irrespective of $f_{\mathrm{m}}$. In BOCDA and BOCDR, an analogous expression, $\Delta z = \Delta\nu_{\mathrm{B}} c/(2\pi n \Delta f f_{\mathrm{m}})$, is well established [13], but the dependence on $f_{\mathrm{m}}$ originates from the finite Brillouin gain bandwidth $\Delta\nu_{\mathrm{B}}$, which has no direct counterpart in reflection-based OCDR. Nevertheless, our recent measurements showed that $\Delta z$ in OCDR with a frequency shifter decreases with increasing $f_{\mathrm{m}}$ [22]. A theoretical description of this behavior and of the parameters governing $\Delta z$ is still lacking.

In this work, we develop a theoretical model for the $f_{\mathrm{m}}$ dependence of the spatial resolution in OCDR with a frequency shifter. The model differs from the conventional treatment in two respects. First, the source linewidth $\delta\nu$ and the ESA resolution bandwidth $B$ are explicitly included, so that the detection is treated as band-limited; the conventional expression corresponds to the limit in which only the $\nu = 0$ modulation sideband is retained. Second, because the experimental signal is measured as electrical power by zero-span detection on an ESA, the spatial response is evaluated from the detected power, with each spectral component weighted by the squared Bessel coefficient. The first point governs the $f_\mathrm{m}$ dependence, whereas the second determines the numerical coefficient of $\Delta z$. The resulting model predicts a low $f_\mathrm{m}$ region in which $\Delta z$ decreases approximately in proportion to $1/f_\mathrm{m}$ and a high $f_\mathrm{m}$ region in which $\Delta z$ approaches a constant value independent of the RBW filter shape. We also measured the RBW filter response and found it to be close to Gaussian, consistent with the response commonly used in commercial spectrum analyzers [23], and incorporated the combined effects of the source linewidth and RBW using a Voigt profile [24]. The model was tested experimentally over a wide range of correlation orders at RBW = 1 MHz. At RBW = 10 MHz, the measured unmodulated beat spectrum was also used directly as the filter function to examine deviations from the analytic model. The modulation amplitude $\Delta f$ was independently calibrated by heterodyne detection [25,26].

# METHODS

## Principle of OCDR with a frequency shifter

In OCDR with a frequency shifter, the output of a laser diode (LD) is sinusoidally frequency modulated and divided into incident and reference beams. The incident beam is launched into the FUT, and the reflected light is combined with the reference beam. An AOM shifts the optical frequency of the reference beam by $f_\mathrm{A}$, so that the heterodyne beat is centered at $f_\mathrm{A}$ rather than near zero frequency, reducing the influence of low-frequency electrical noise. The beat signal is detected by a photodetector and measured by an ESA in zero-span mode at a center frequency of $f_\mathrm{A}$ with resolution bandwidth $B$. The following analysis derives the detected electrical power and the corresponding spatial resolution from the beat spectrum [18,19].

## Theoretical formulation of spatial resolution

### Beat spectrum of sinusoidally frequency-modulated light

The instantaneous optical frequency of the sinusoidally modulated light source is

$$f(t) = f_0 + \Delta f \cdot \sin(2\pi f_\mathrm{m} t) \tag{1}$$

Integrating Eq. (1) gives the optical phase. For light reflected from a point at a distance $x$ from a correlation peak, the relative delay is $\tau = 2nx/c$. After the frequency shift by $f_\mathrm{A}$, the heterodyne beat can be expanded using the Jacobi-Anger expansion [18,19]. The resulting spectrum consists of discrete components at $f_\mathrm{A} + \nu f_\mathrm{m}$, where $\nu$ is an integer. The complex amplitude of the $\nu$-th component is proportional to $J_\nu[\beta(x)]$, where $J_\nu$ is the Bessel function of the first kind of order $\nu$ and

$$\beta(x) = 2\frac{\Delta f}{f_\mathrm{m}}\left|\sin\left(2\pi n f_\mathrm{m}\frac{x}{c}\right)\right| \tag{2}$$

Correlation peaks occur at $\beta(x) = 0$ and therefore recur at

$$x_\mathrm{m} = m \cdot \frac{c}{2nf_\mathrm{m}}, \qquad m = 0, 1, 2, \dots. \tag{3}$$

The spacing between adjacent correlation peaks, $c/(2nf_\mathrm{m})$, gives the unambiguous measurement range [8,9]. For a fixed reflector at a distance $x_0$ from the reference point, the $m$-th correlation peak coincides with the reflector when $f_{\mathrm{m}} = m\, f_{\mathrm{m1}}$, where $f_{\mathrm{m1}} = c/(2nx_0)$. This relation is used below to define the correlation order $m$.

**Electrical power spectrum observed by the ESA**

The photodetector converts the heterodyne optical signal into a photocurrent, and the ESA measures its electrical power spectrum. Because the complex amplitude of the $\nu$-th beat component is proportional to $J_\nu[\beta(x)]$, its electrical power is proportional to $J_\nu^2[\beta(x)]$. The normalized beat power spectrum is therefore

$$S_{\mathrm{b}}(x,f) = \sum_{\nu=-\infty}^{\infty} J_\nu^2[\beta(x)] \cdot \delta(f - f_{\mathrm{A}} - \nu f_{\mathrm{m}}) \ . \tag{4}$$

The identity $\Sigma_\nu J_\nu^2(\beta) = 1$ indicates that the total beat power is conserved as it is redistributed among the spectral components. At a correlation peak, $\beta = 0$ and all the power is concentrated in the $\nu = 0$ component. Away from the peak, the power is distributed among multiple components.

Conventional SOCF theory defines the spatial resolution from the half-width of the synthesized optical coherence function, which corresponds to the visibility of the interference fringes [8,9,15] and hence to the beat-current amplitude [18]. This gives $\Delta z = 0.76c/(\pi n \Delta f)$ from $|J_0(\beta)| = 1/2$. In the present system, however, $\Delta z$ is obtained from the electrical power measured by the ESA. We therefore define the spatial response using $J_\nu^2$ rather than $J_\nu$. This definition reflects the measured observable and does not modify the conventional definition when the beat-current amplitude itself is measured.

**Filter function including the source linewidth and ESA RBW**

Equation (4) assumes infinitely narrow spectral components. In practice, the finite source linewidth and the ESA RBW must be considered. The source line shape is modeled as a Lorentzian with FWHM $\delta\nu$:

$$L(f) = \frac{1}{\pi} \cdot \frac{\gamma}{f^2 + \gamma^2}, \qquad \gamma = \frac{\delta\nu}{2} \ . \tag{5}$$

The ESA passes each broadened component through its RBW filter centered at $f_{\mathrm{A}}$. As shown in Results, the measured power transmission profile of the RBW filter is close to Gaussian, with FWHM $B$, consistent with the typical response of commercial spectrum analyzers [23]:

$$H(f) = \exp\left(-4 \ln 2 \cdot \frac{f^2}{B^2}\right) , \tag{6}$$

where $f$ is the frequency offset from the filter center. The fraction of the $\nu$-th spectral component transmitted through the filter is then

$$F_\nu = \int_{-\infty}^{\infty} L(f - \nu f_{\mathrm{m}}) \cdot H(f) df \ . \tag{7}$$

Equation (7) is a Voigt profile and can be evaluated using the Faddeeva function $w(z) =$

$\exp(-z^2)\cdot \mathrm{erfc}(-jz)$ [24]:

$$F_\nu \propto \mathrm{Re}[w(z_\nu)], \qquad z_\nu = \frac{\nu f_\mathrm{m} + j\gamma}{\sigma\sqrt{2}}, \qquad \sigma = \frac{B}{2\sqrt{2\ln 2}}\,. \tag{8}$$

Because only the ratio $F_\nu/F_0$ is used below, the normalization of $F_\nu$ does not affect the calculated spatial resolution.

**Full numerical model and spatial resolution**

The electrical power observed by the ESA at position $x$ is obtained by summing the transmitted power of all spectral components:

$$P(x) = \sum_{\nu=-\infty}^{\infty} J_\nu^2[\beta(x)]\, F_\nu\,. \tag{9}$$

At a correlation peak, $\beta = 0$ and only the $\nu = 0$ component contributes, giving $P(0) = F_0$. We define the spatial resolution $\Delta z$ as the FWHM of $P(x)$. Thus, if $x_{1/2}$ is the first positive position satisfying $P(x_{1/2})/P(0) = 1/2$, then $\Delta z = 2x_{1/2}$. Equation (9) is the full numerical model used in this study. In general, $x_{1/2}$ must be obtained numerically because both $J_\nu^2[\beta(x)]$ and $F_\nu$ depend on the system parameters. Closed-form expressions can, however, be derived in the limits of large and small $f_m$, as described below.

**Limit I: Saturation region at high $f_\mathrm{m}$**

When $f_\mathrm{m}$ is much larger than both $B$ and $\delta\nu$, the spacing between adjacent beat components exceeds the width of the filter function. Thus, $F_\nu \approx 0$ for $\nu \neq 0$, and Eq. (9) reduces to $P(x) \approx J_\nu^2[\beta(x)]\, F_0$. Because $P(0) = F_0$, the normalized spatial response is

$$\frac{P(x)}{P(0)} = J_0^2[\beta(x)]\,. \tag{10}$$

The half-power condition is therefore $J_0^2(\beta_{1/2}) = 1/2$, where $\beta_{1/2}$ denotes the value of $\beta$ at $x = x_{1/2}$. Solving this equation gives $\beta_{1/2}$ = 1.126. Using the small-angle approximation in Eq. (2), $\beta(x) \approx (4\pi n\Delta f/c)\,x$, the spatial resolution becomes

$$\Delta z_\mathrm{sat} = \frac{0.563c}{\pi n\Delta f}\,. \tag{11}$$

Equation (11) contains neither $B$ nor $\delta\nu$. In this limit, only the $\nu = 0$ component contributes to the detected power, and the factor $F_0$ cancels upon normalization. The saturation resolution is therefore independent of the RBW filter shape, the RBW, and the source linewidth. For comparison, the conventional SOCF definition based on the fringe visibility or equivalently the beat-current amplitude gives $\Delta z = 0.76c/(\pi n\Delta f)$ from $|J_0(\beta)| = 1/2$ [9,18]. The difference between the two coefficients results from evaluating electrical power rather than beat-current amplitude.

**Limit II: Tail region at low $f_\mathrm{m}$**

When $f_\mathrm{m}$ is much smaller than the width of the filter function, many spectral components contribute to Eq. (9), and the discrete comb can be approximated as a continuum. The beat-frequency deviation spans the range from $-W(x)$ to $W(x)$, where

$$W(x) = \beta(x)\, f_{\mathrm{m}} = 2\Delta f \left|\sin\left(2\pi n f_{\mathrm{m}} \frac{x}{c}\right)\right| . \tag{12}$$

For sinusoidal frequency modulation, the normalized distribution of the instantaneous frequency deviation is

$$\rho_W(f, x) = \frac{1}{\pi\sqrt{W^2(x) - f^2}}, \qquad |f| < W(x). \tag{13}$$

Let $F(f)$ denote the continuous filter function corresponding to Eq. (7). Replacing the discrete sum in Eq. (9) by an integral gives

$$P_{\mathrm{cont}}(x) = \int_{-W(x)}^{W(x)} \rho_W(f)\, F(f)\, df . \tag{14}$$

Using $f = W(x)\sin\theta$, Eq. (14) becomes

$$P_{\mathrm{cont}}(x) = \frac{1}{\pi}\int_{-\frac{\pi}{2}}^{\frac{\pi}{2}} F(W(x)\sin\theta) d\theta . \tag{15}$$

We define $W_{1/2} = W(x_{1/2})$, where $x_{1/2}$ satisfies $P_{\mathrm{cont}}(x_{1/2}) = P_{\mathrm{cont}}(0)/2$. The corresponding spatial resolution follows from Eq. (12). In the tail region, where the half-power point is sufficiently close to the correlation peak, the small-angle approximation gives $\Delta z_{\mathrm{tail}} = cW_{1/2}/(2\pi n \Delta f f_{\mathrm{m}})$. The value of $W_{1/2}$ depends on both $\delta\nu$ and $B$. Closed-form expressions can be obtained when either the RBW or the source linewidth dominates, as described below.

**Limit II-a: $\delta\nu \ll B$ (Gaussian-dominated filter function)**

When the source linewidth is negligible compared with the RBW, the Lorentzian contribution to $F$ approaches a delta function, and the continuous filter function reduces to the Gaussian function $H(f)$ in Eq. (6). Substituting $F(f) = H(f)$ into Eq. (15) gives

$$P_{\mathrm{cont}}(x) = \frac{1}{\pi}\int_{-\frac{\pi}{2}}^{\frac{\pi}{2}} \exp\left(-4\ln 2 \cdot \frac{W^2(x)\sin^2\theta}{B^2}\right) d\theta . \tag{16}$$

Using $\sin^2\theta = (1 - \cos 2\theta)/2$ and defining $a(x) = 2\ln 2 \cdot W^2(x)/B^2$, Eq. (16) becomes

$$P_{\mathrm{cont}}(x) = e^{-a(x)} \cdot \frac{1}{\pi}\int_{-\frac{\pi}{2}}^{\frac{\pi}{2}} \exp(a(x)\cos 2\theta)\, d\theta . \tag{17}$$

With the substitution $u = 2\theta$,

$$P_{\mathrm{cont}}(x) = e^{-a(x)} \cdot \frac{1}{2\pi} \int_{-\pi}^{\pi} \exp(a(x) \cos u)\, du = e^{-a(x)} \cdot I_0\big(a(x)\big)\,, \tag{18}$$

where $I_0$ is the modified Bessel function of the first kind of order zero. At the correlation peak, $W(0) = 0$ and hence $P_{\mathrm{cont}}(0) = 1$. At the half-power point $x = x_{1/2}$, we define $a_{1/2} = 2 \ln 2 \cdot W_{1/2}^2 / B^2$. The half-power condition is therefore $e^{-a_{1/2}} \cdot I_0\big(a_{1/2}\big) = 1/2$, which gives $a_{1/2} = 0.8768$. Thus,

$$W_{1/2}|_{\delta\nu \ll B} = B \sqrt{\frac{a_{1/2}}{2 \ln 2}} \approx 0.7953B\,. \tag{19}$$

For an ideal rectangular filter with the same FWHM, $W_{1/2} = B/\sqrt{2} = 0.7071B$. The Gaussian value is approximately 12% larger because the Gaussian filter transmits finite power outside its FWHM.

**Limit II-b: $\delta\nu \gg B$ (Lorentzian-dominated filter function)**

When the source linewidth is much larger than the RBW, the Lorentzian varies little over the Gaussian passband. Thus, $L(f' - f) \approx L(f)$ within the passband, and the filter function can be written as:

$$F(f) \approx L(f) \cdot \int_{-\infty}^{\infty} H(f')df' = L(f) \cdot A, \qquad A = B\sqrt{\frac{\pi}{4 \ln 2}}\;. \tag{20}$$

Substituting Eq. (20) and the arcsine distribution in Eq. (13) into Eq. (14), with $L(f) = (\gamma/\pi)/(f^2 + \gamma^2)$, gives

$$P_{\mathrm{cont}}(x) = \frac{A\gamma}{\pi^2} \cdot \int_{-W(x)}^{W(x)} \frac{df}{(f^2 + \gamma^2)\sqrt{W^2(x) - f^2}}\;. \tag{21}$$

Using $f = W(x) \sin\theta$,

$$P_{\mathrm{cont}}(x) = \frac{A\gamma}{\pi^2} \int_{-\frac{\pi}{2}}^{\frac{\pi}{2}} \frac{d\theta}{W^2(x) \sin^2\theta + \gamma^2}\;. \tag{22}$$

With the substitution $u = \tan\theta$,

$$P_{\mathrm{cont}}(x) = \frac{A\gamma}{\pi^2} \int_{-\infty}^{\infty} \frac{du}{(W^2(x) + \gamma^2) \cdot u^2 + \gamma^2}\;. \tag{23}$$

Evaluating the integral gives

$$P_{\mathrm{cont}}(x) = \frac{A}{\pi\sqrt{W^2(x) + \gamma^2}}\;. \tag{24}$$

At the correlation peak, $W(0) = 0$ and $P_{\mathrm{cont}}(0) = A/(\pi\gamma)$. The half-power condition $P_{\mathrm{cont}}(x_{1/2}) = P_{\mathrm{cont}}(0)/2$ therefore gives

$$W_{1/2}|_{\delta\nu\gg B} = \sqrt{3}\gamma = \frac{\sqrt{3}}{2}\delta\nu \approx 0.866\delta\nu\,. \tag{25}$$

In this limit, only the area $A$ of the RBW filter enters the detected power, and $A$ cancels in the half-power condition. Thus, $W_{1/2}$ is independent of the RBW filter shape.

### Interpolation between the sub-limits and the tail-region formula

For intermediate values of $\delta\nu/B$, $W_{1/2}$ must be obtained numerically from Eqs. (14) and (15) using the full Voigt filter function. For practical use, the two limiting values in Eqs. (19) and (25) can be approximated by the power-mean interpolation

$$W_{1/2} \approx [(0.7953B)^p + (0.866\delta\nu)^p]^{\frac{1}{p}}\,. \tag{26}$$

Numerical optimization gives $p = 1.4$, for which Eq. (26) agrees with the numerical solution within 3.62% over the range of $\delta\nu/B$ considered here (see Numerical procedures). In the tail region, $x_{1/2}$ is sufficiently small that Eq. (12) can be approximated as $W(x) \approx (4\pi n\Delta f f_{\mathrm{m}}/c)x$. Setting $W(x_{1/2}) = W_{1/2}$ and $\Delta z = 2x_{1/2}$ gives

$$\Delta z_{\mathrm{tail}} = \frac{cW_{1/2}}{2\pi n\Delta f f_{\mathrm{m}}}\,. \tag{27}$$

Equation (27) has the same $f_{\mathrm{m}}$ dependence as the BOCDA and BOCDR expression $\Delta z = c\,\Delta\nu_{\mathrm{B}}/(2\pi n\Delta f f_{\mathrm{m}})$ [13], with the Brillouin gain bandwidth $\Delta\nu_{\mathrm{B}}$ replaced by $W_{1/2}$, which accounts for the source linewidth and ESA RBW. The saturation and tail expressions, Eqs. (11) and (27), can then be combined as

$$\Delta z_{\mathrm{interp}} = \left[\Delta z_{\mathrm{sat}}^q + \Delta z_{\mathrm{tail}}^q\right]^{\frac{1}{q}}, \tag{28}$$

where $q = 3.5$ is used as a representative value based on numerical optimization against Eq. (9) (see Numerical procedures). The two asymptotes are equal at $f_{\mathrm{m}}^* = W_{1/2}/1.126$, which provides an approximate boundary between the tail and saturation regions.

## Model-free filter function from the unmodulated beat spectrum

The analytic filter function in Eq. (8) requires the source linewidth $\delta\nu$ and the RBW $B$ to be specified and assumes Lorentzian and Gaussian line shapes, respectively. To avoid these assumptions, we also obtained the filter function directly from the measured unmodulated beat spectrum. With the sinusoidal frequency modulation switched off, the measured spectrum represents the convolution of the source line shape and the actual RBW filter response under the same measurement conditions.

The measured spectrum was baseline subtracted using the median of the lowest 10% of the measured values, normalized to its peak value, and linearly interpolated to obtain the continuous filter function $F(f)$. For the full numerical model, $F_\nu$ in Eq. (9) was obtained by evaluating $F(f)$ at $f = \nu f_{\mathrm{m}}$. The same measured $F(f)$ was also substituted into Eq. (15), and $W_{1/2}$ was determined numerically from the condition $P_{\mathrm{cont}}(x_{1/2}) = P_{\mathrm{cont}}(0)/2$. This measured value of $W_{1/2}$ was then used in Eq. (27).

This approach does not assume a specific source line shape or RBW filter shape, while retaining the full numerical model in Eq. (9) and the tail-region relation in Eq. (27).

## Numerical procedures

The constants $\beta_{1/2}$ and $a_{1/2}$ were obtained numerically from $J_0^2(\beta_{1/2}) = 1/2$ and $e^{-a_{1/2}} I_0(a_{1/2}) = 1/2$, respectively, using Newton's method. The solutions were $\beta_{1/2} = 1.126364$ and $a_{1/2} = 0.876842$.

The interpolation exponent $p$ in Eq. (26) was determined by comparing the interpolated and numerical values of $W_{1/2}$. The ratio $r = \delta\nu/B$ was varied logarithmically from $10^{-2}$ to $10^{2}$. For each $r$, $W_{1/2}$ was obtained from Eqs. (14) and (15) using the Voigt filter function and Brent's method. Candidate values of $p$ from 1.0 to 2.5 in increments of 0.05 were evaluated. The minimum maximum relative deviation was obtained at $p = 1.4$, with a maximum deviation of 3.62% over the tested range (Table 1). The exponent $q$ in Eq. (28) was similarly optimized against the numerical solution of Eq. (9) over the experimentally investigated range of $f_{\mathrm{m}}$. The optimal value varied with RBW, as summarized in Table 2. A common value of $q = 3.5$ was used for all calculations and kept the maximum relative deviation below approximately 6.3% for all tested RBW settings.

**Table 1.** Maximum relative deviation of Eq. (26) from the numerically obtained $W_{1/2}$ over $\delta\nu/B \in [10^{-2}, 10^{2}]$ for representative values of $p$.

| $p$ | Max relative deviation [%] |
|---|---|
| 1.0 | 22.78 |
| 1.2 | 10.20 |
| 1.4 | 3.62 |
| 1.6 | 7.88 |
| 2.0 | 14.71 |

**Table 2.** Optimal interpolation exponent $q$ in Eq. (28) for each RBW setting, obtained by minimizing the maximum relative deviation from the numerical solution of Eq. (9) over the experimentally relevant $f_{\mathrm{m}}$ range. A common value $q = 3.5$ was used throughout this study.

| RBW | Optimal $q$ | Max relative deviation [%] |
|---|---|---|
| 30 kHz | 3.50 | 4.52 |
| 100 kHz | 3.60 | 4.72 |
| 300 kHz | 3.85 | 5.28 |
| 1 MHz | 4.10 | 5.03 |
| 3 MHz | 3.40 | 1.27 |
| 5 MHz | 3.05 | 2.78 |
| 10 MHz | 3.05 | 3.69 |

All theoretical curves were calculated using Python with NumPy and SciPy. For each $f_{\mathrm{m}}$, $x$ was uniformly sampled up to $x_{\mathrm{max}}$, which was set to four times an approximate value of $\Delta z$ estimated from Eqs. (11) and (27) by quadrature. Typically, 400 to 600 points were used. The summation over $\nu$ in Eq. (9) was truncated at $|\nu| \leq \nu_{\mathrm{max}}$, where $\nu_{\mathrm{max}}$ was chosen to include both the range of significant Bessel components and the width of the filter function:

$$\nu_{\max} = \min\left\{4000, \left\lceil \max_x[\beta(x)] + 10\sqrt{\max_x[\beta(x)] + 1 + \frac{B/2 + 10\gamma}{f_{\mathrm{m}}}} \right\rceil\right\}. \quad (29)$$

For each $x, P(x)$ was evaluated from Eq. (9) using scipy.special.jv for the Bessel functions and scipy.special.wofz for the Faddeeva function. For the measured filter function, linear interpolation of the measured spectrum was used instead. The first half-power crossing was identified from $P(x)/P(0) = 1/2$ and refined by linear interpolation between the adjacent sampled points. The spatial resolution was then calculated as $\Delta z = 2x_{1/2}$. The integral in Eq. (15) was evaluated over $-\pi/2 \leq \theta \leq \pi/2$ using fixed-node trapezoidal integration. For the measured filter function, 2001 nodes were used to avoid instability caused by adaptive integration of the linearly interpolated measured spectrum. Finally, $W_{1/2}$ was obtained from $P_{\mathrm{cont}}(x_{1/2}) = P_{\mathrm{cont}}(0)/2$ using Brent's method. Table 3 summarizes the key equations used to calculate the spatial resolution.

**Table 3.** Summary of the key equations used to calculate the spatial resolution.

| Eq. | Expression | Description |
|---|---|---|
| 9 | $P(x) = \sum_{\nu=-\infty}^{\infty} J_\nu^2[\beta(x)] \cdot F_\nu$ | Full numerical model for the observed electrical power |
| 11 | $\Delta z_{\mathrm{sat}} = \frac{0.563c}{\pi n \Delta f}$ | Spatial resolution in the high $f_{\mathrm{m}}$ saturation region |
| 26 | $W_{1/2}(B, \delta\nu) \approx [(0.7953B)^p + (0.866\delta\nu)^p]^{\frac{1}{p}}$ | Beat-frequency sweep amplitude at the half-power point in the continuous-limit response; $p = 1.4$ |
| 27 | $\Delta z_{\mathrm{tail}} = \frac{cW_{1/2}(B, \delta\nu)}{2\pi n \Delta f f_{\mathrm{m}}}$ | Spatial resolution in the low $f_{\mathrm{m}}$ tail region |
| 28 | $\Delta z_{\mathrm{interp}} = \left[\Delta z_{\mathrm{sat}}^q + \Delta z_{\mathrm{tail}}^q\right]^{\frac{1}{q}}$ | Interpolated spatial resolution; $q =$ 3.5 |
| — | $f_{\mathrm{m}}^* = \frac{W_{1/2}(B, \delta\nu)}{\beta_{1/2}} \approx \frac{W_{1/2}(B, \delta\nu)}{1.126}$ | Crossover modulation frequency |

## Experimental setup and measurement procedure

The experimental setup is shown in Fig. 1. A distributed-feedback LD (LD1, nominal linewidth ~2 MHz) was directly frequency-modulated by a function generator at $f_{\mathrm{m}}$ with a modulation amplitude $\Delta f \approx 5.0$ GHz. The value of $\Delta f$ was calibrated at each $f_{\mathrm{m}}$ as described below. LD1 and a second unmodulated LD (LD2) were temperature-stabilized using thermoelectric coolers (TECs) to maintain their optical frequency difference within the bandwidth of the heterodyne photodetector (PD2) used for $\Delta f$ calibration. The LD1 output was divided at a 90:10 fiber coupler. The 90% branch was further divided into incident and reference beams for OCDR. The incident beam passed through a 10 km delay fiber and an erbium-doped fiber amplifier (EDFA1, output power ≈ 10.0 dBm) before being launched through an optical circulator into a 2 m FUT terminated with a physical-contact (PC) open end. The reference beam was frequency-shifted by $f_{\mathrm{A}} = 80$ MHz using the AOM, with an output power of approximately +1.3 dBm after the AOM. Light reflected from the FUT was amplified by EDFA2 (output power ≈ −16.7 dBm) and combined with the reference beam before photodetector PD1. The resulting heterodyne beat was measured by ESA1 (Advantest R3267) in zero-span mode at a center frequency of $f_{\mathrm{A}}$.

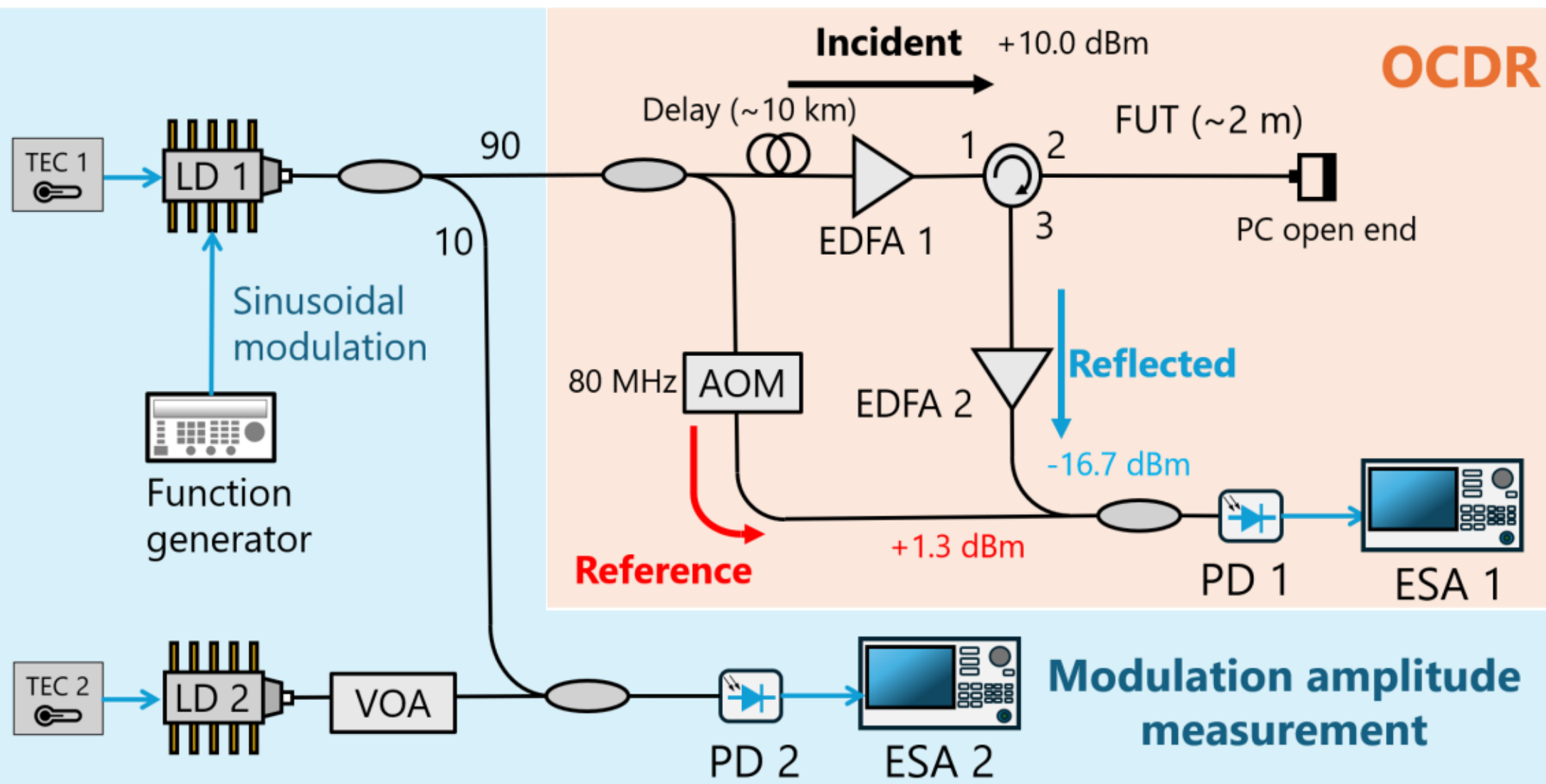

**Fig. 1.** Experimental setup for OCDR with a frequency shifter and heterodyne calibration of the modulation amplitude $\Delta f$. AOM: acousto-optic modulator, EDFA: erbium doped fiber amplifier, ESA: electrical spectrum analyzer, FUT: fiber under test, LD: laser diode, PD: photodetector, TEC: thermoelectric cooler, VOA: variable optical attenuator.

The 10% branch of LD1 was combined with the output of LD2 after attenuation by a variable optical attenuator (VOA) and detected by PD2 and ESA2. Because LD2 was unmodulated, the heterodyne spectrum reproduced the instantaneous frequency excursion of LD1, yielding an arcsine distribution extending from $-\Delta f$ to $+\Delta f$. The value of $\Delta f$ was determined from the outer edges of this distribution, where the measured power fell by 3 dB from the corresponding edge peaks.

OCDR measurements were performed at the $m$-th correlation order, defined by $f_{\mathrm{m}} = m\, f_{\mathrm{m1}}$, where $f_{\mathrm{m1}} \approx 20.24$ kHz is the fundamental modulation frequency at which the first correlation peak coincides with the fixed FUT reflector. For each $m, f_{\mathrm{m}}$ was swept over a narrow range around $m f_{\mathrm{m1}}$ while ESA1 recorded the power at the fixed center frequency $f_{\mathrm{A}}$. Near a fixed reflector, varying $f_{\mathrm{m}}$ shifts the correlation peak relative to the reflector through $\beta(x)$, allowing the spatial response $P(x)/P(0)$ to be obtained without physically scanning the reflector position. ESA1 was operated with RBW = 1 or 10 MHz, video bandwidth (VBW) = 30 kHz, and a zero-span sweep time of 40 ms. Each trace was averaged over 16 sweeps. The half-power points were determined after baseline subtraction, and the corresponding modulation-frequency offsets were converted to spatial displacements using $|dx/df_{\mathrm{m}}| \approx c/(2n f_{\mathrm{m1}} f_{\mathrm{m}})$. The spatial resolution $\Delta z$ was then obtained from the separation between the two half-power points.

The RBW filter response of ESA1 was characterized independently by applying an 80 MHz continuous-wave sinusoid from the function generator directly to the ESA and sweeping the ESA center frequency across the tone. Measurements were performed for RBWs from 30 kHz to 10 MHz. Because the linewidth of the reference tone was negligible compared with the tested RBWs, the measured trace represented the power transmission profile $H(f)$ of the RBW filter. The unmodulated beat spectrum used as the measured filter function was obtained with the sinusoidal frequency modulation of LD1 switched off. The resulting spectrum represents the source line shape convolved with the RBW filter response under the actual optical measurement conditions. The source linewidth used in the calculations, $\delta\nu = 2.57$ MHz, was obtained from the FWHM of the unmodulated beat spectrum measured at RBW = 30 kHz, for which the RBW contribution to the measured linewidth was negligible.

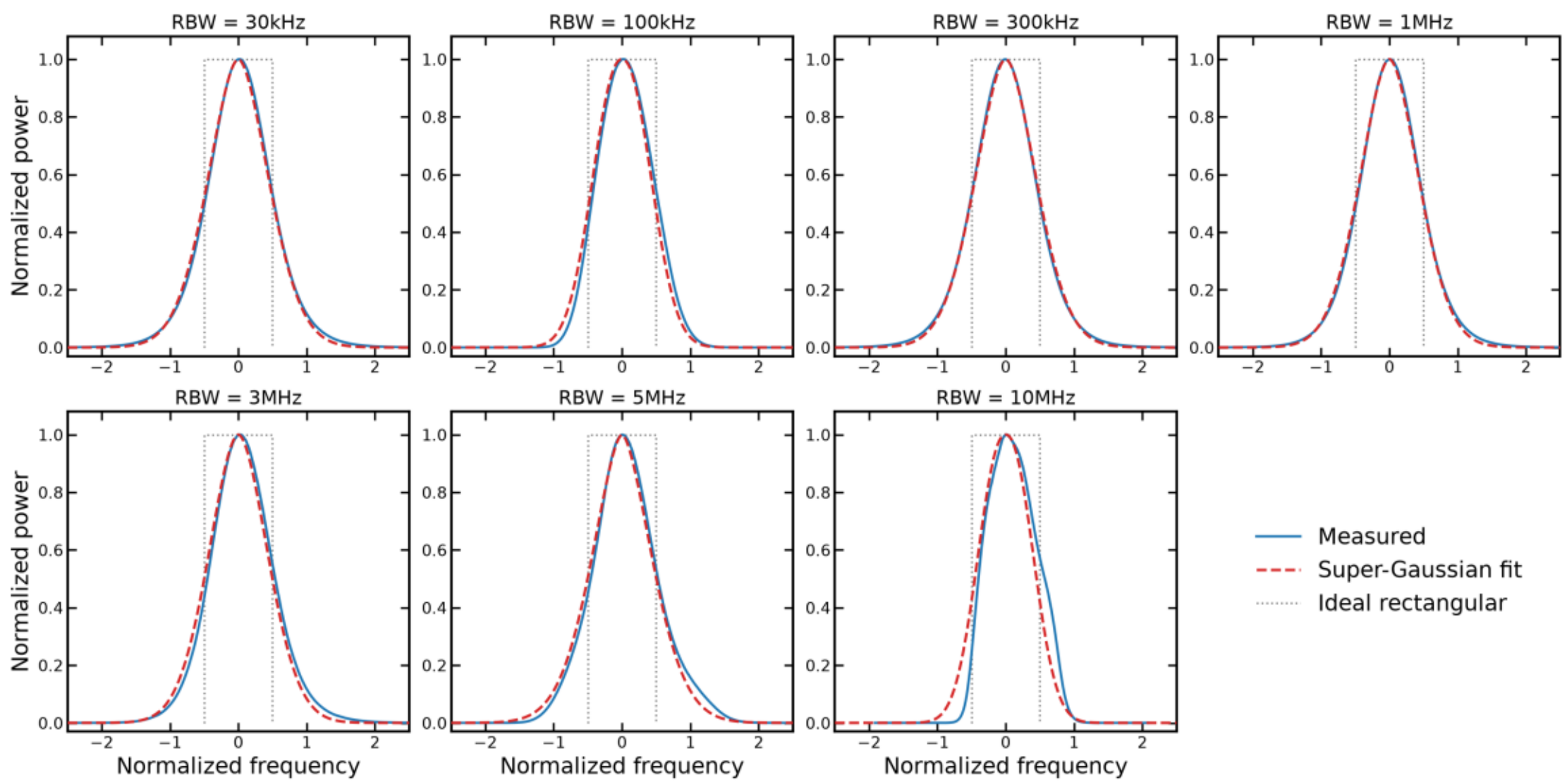

**Fig. 2.** Measured RBW filter profiles of the ESA for RBW = 30 kHz, 100 kHz, 300 kHz, 1 MHz, 3 MHz, 5 MHz, and 10 MHz. Solid lines show measurements with a continuous-wave reference tone, dashed lines show super-Gaussian fits, and dotted lines indicate an ideal rectangular filter. The horizontal axis is the frequency offset normalized by the RBW setting.

## RESULTS

### Experimental characterization of the RBW filter shape

The measured RBW filter profiles are shown in Fig. 2. For RBW settings from 30 kHz to 5 MHz, the measured FWHM agreed with the nominal values within approximately 3%. At RBW = 10 MHz, the measured FWHM was 9.05 MHz, approximately 9.5% lower than the nominal value. Fitting each profile with the super-Gaussian function $\exp[-\ln 2 \cdot |2f/W|^{2k}]$ yielded $k = 0.84$ to 1.11, where $k = 1$ corresponds to a Gaussian profile. The coefficient of determination was $R^2 = 0.99$ for all RBW settings except 10 MHz, for which $R^2 = 0.95$. These results show that the measured RBW filter profiles were close to Gaussian over the tested RBW range.

### Spatial resolution at RBW = 1 MHz

Figure 3 shows the measured spatial resolution $\Delta z$ as a function of the modulation frequency $f_{\mathrm{m}}$ at RBW = 1 MHz. Measurements were performed for correlation orders $m = 1$ to 2048, corresponding to $f_{\mathrm{m}} \approx 20$ kHz to 41 MHz, with 10 repeated measurements at each order. At low $f_{\mathrm{m}}$, $\Delta z$ decreased approximately in proportion to $1/f_{\mathrm{m}}$, as predicted by Eq. (27). The decrease became weaker around $f_{\mathrm{m}} = 1$ to 2 MHz, and $\Delta z$ approached approximately 0.7 to 0.8 cm at high $f_m$. The full numerical model in Eq. (9), calculated using RBW = 1 MHz and $\delta\nu = 2.57$ MHz, reproduced the overall dependence with a root mean square relative deviation of 13.6%. The mean relative deviation was −4.4% at low correlation orders and +2.9% at high correlation orders. The largest deviations occurred at the lowest orders, where the calculated values exceeded the measured values by approximately 20 to 29%. At high $f_m$, the measured $\Delta z$ approached the saturation value predicted by Eq. (11).

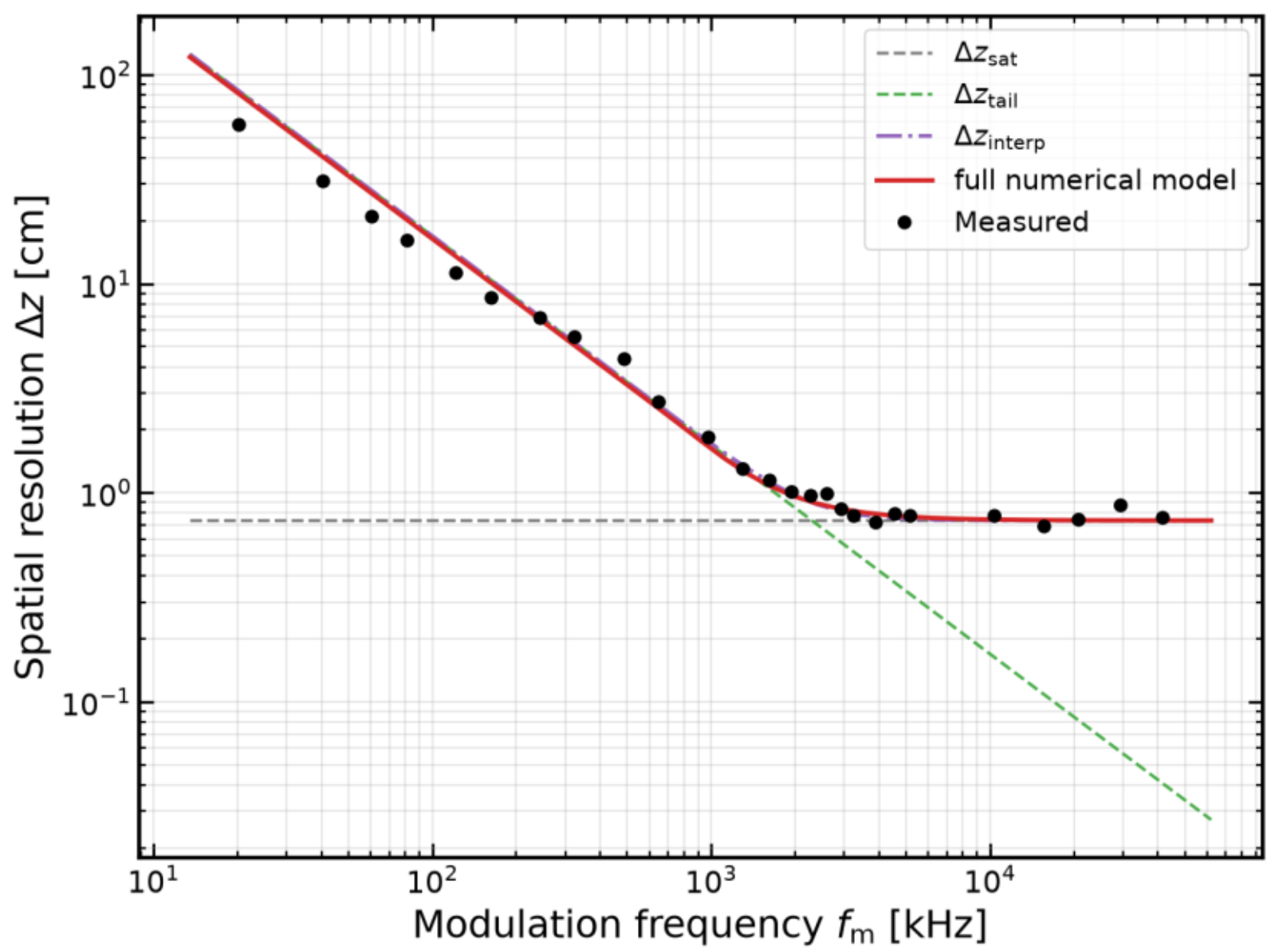

**Fig. 3.** Spatial resolution $\Delta z$ as a function of modulation frequency $f_m$ at RBW = 1 MHz. Points show the mean of 10 repeated measurements at each of 26 correlation orders spanning $m = 1$ to 2048. Curves show the saturation-region asymptote $\Delta z_{sat}$ [Eq. (11)], tail-region asymptote $\Delta z_{tail}$ [Eqs. (26) and (27)], power-mean interpolation $\Delta z_{interp}$ [Eq. (28)], and full numerical model [Eq. (9)].

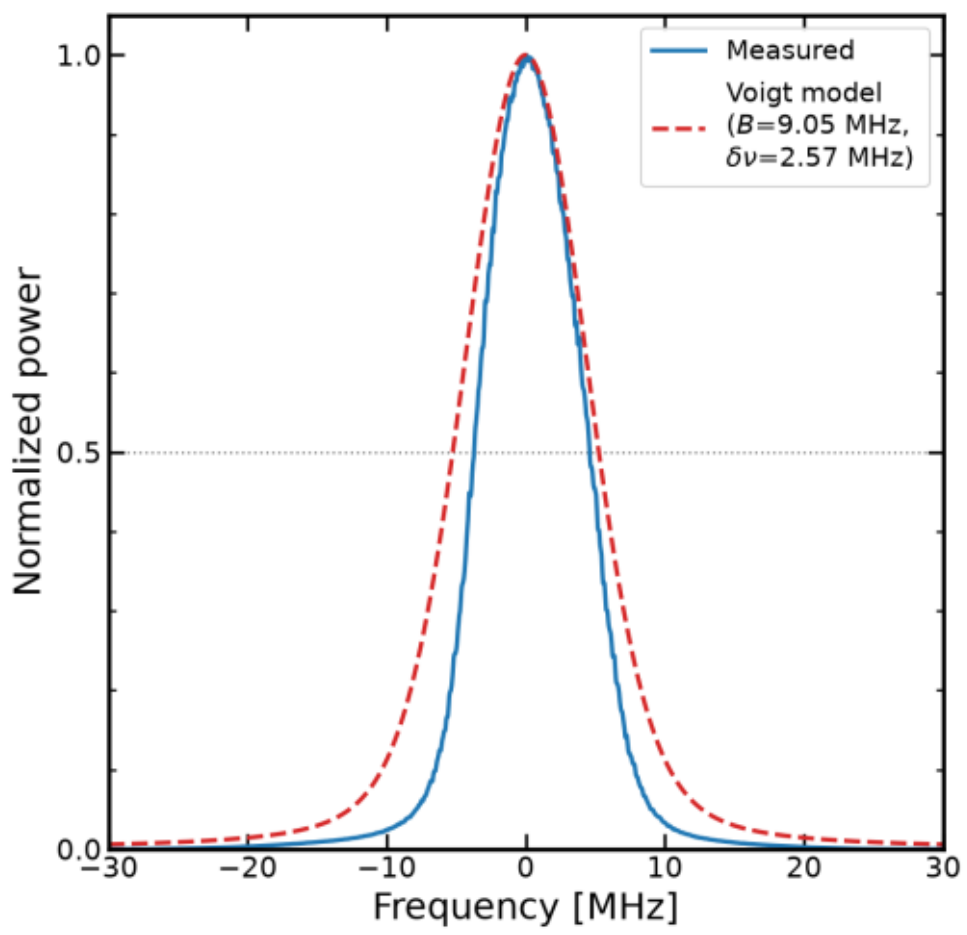

**Fig. 4.** Measured filter function at RBW = 10 MHz. The baseline-subtracted unmodulated beat spectrum, normalized to unit peak, is compared with the Voigt model in Eq. (8), calculated using the independently measured RBW filter FWHM $B = 9.05$ MHz and source linewidth $\delta\nu = 2.57$ MHz.

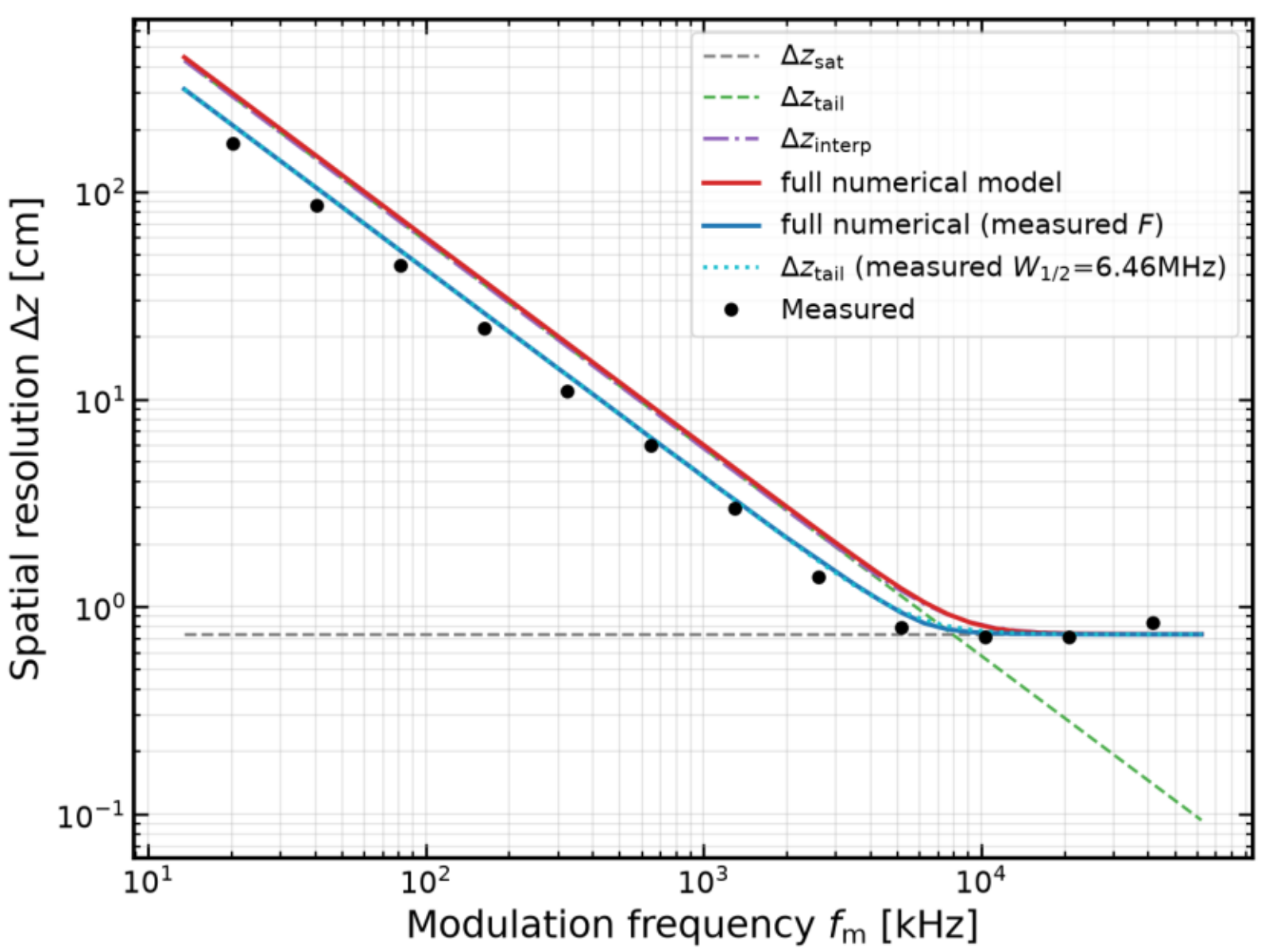

**Fig. 5.** Spatial resolution $\Delta z$ as a function of modulation frequency $f_m$ at RBW = 10 MHz. Points show measured values at each of 12 correlation orders spanning $m = 1$ to 2048.. Curves show the saturation-region asymptote $\Delta z_{sat}$ [Eq. (11)], tail-region asymptote $\Delta z_{tail}$ [Eqs. (26) and (27)], power-mean interpolation $\Delta z_{interp}$ [Eq. (28)], the full numerical model [Eq. (9)] using the Voigt filter function in Eq. (8), the full numerical model using the measured unmodulated beat spectrum directly as $F(f)$, and Eq. (27) evaluated with the measured $W_{1/2}$ = 6.46 MHz. The Voigt-based calculations use $B = 9.05$ MHz and $\delta\nu = 2.57$ MHz.

### Measured filter function at RBW = 10 MHz

Figure 4 compares the unmodulated beat spectrum measured at RBW = 10 MHz with the Voigt filter function in Eq. (8). The Voigt profile was calculated using the independently measured RBW filter FWHM, $B = 9.05$ MHz, and source linewidth, $\delta\nu = 2.57$ MHz. The measured spectrum was narrower than the Voigt profile over most of the measured frequency range. The measured value was $W_{1/2} = 6.46$ MHz, approximately 24% smaller than the Voigt-model value of 8.47 MHz. This difference was subsequently examined by using the measured spectrum directly as $F(f)$ in Eq. (9).

### Spatial resolution at RBW = 10 MHz

Figure 5 shows the measured spatial resolution $\Delta z$ as a function of $f_m$ at RBW = 10 MHz. For correlation orders $m = 1$ to 128, the full numerical model in Eq. (9), using the Voigt filter function with $B$ = 9.05 MHz and $\delta\nu = 2.57$ MHz, systematically overestimated $\Delta z$. The measured values were 57 to 64% of the calculated values. At high $f_m$, however, the measured $\Delta z$ approached the saturation value in Eq. (11), with a difference of approximately 3%.

Replacing the Voigt filter function in Eq. (9) with the measured unmodulated beat spectrum substantially reduced the discrepancy. Over $m = 1$ to 2048, the measured $\Delta z$ values were 81 to 91% of those calculated using the measured filter function. The tail-region expression in Eq. (27), evaluated using $W_{1/2}$ obtained from the measured spectrum, showed a similar dependence on $f_m$.

## DISCUSSION

The present results identify two main features of the spatial resolution in OCDR with a frequency shifter. First, $\Delta z$ decreases with increasing $f_{\mathrm{m}}$ in the low $f_{\mathrm{m}}$ region and approaches a constant value at high $f_{\mathrm{m}}$. This behavior is reproduced by the power-based model that includes the source linewidth $\delta\nu$ and ESA RBW $B$, and the measured high $f_{\mathrm{m}}$ limit agrees with the saturation value in Eq. (11). Second, the accuracy of the calculated $\Delta z$ depends on how the combined source and receiver response is represented. At RBW = 10 MHz, the Voigt model overestimated $\Delta z$, whereas using the measured unmodulated beat spectrum as $F(f)$ reduced the discrepancy. These results show that both the power-detection scheme and the effective spectral response of the measurement system must be considered when predicting the spatial resolution of OCDR.

The observed $f_{\mathrm{m}}$ dependence cannot be described by the conventional SOCF expression, $\Delta z = 0.76c/(\pi n \Delta f)$, which is independent of $f_{\mathrm{m}}$ [9,18]. The conventional expression is obtained from a time average of the interference signal, which retains only the $\nu = 0$ component of the beat spectrum and therefore corresponds to a vanishing detection bandwidth. In the present model, this dependence arises because the ESA detects the signal within a finite bandwidth. At low $f_{\mathrm{m}}$, multiple spectral components fall within the detection bandwidth, and the spatial resolution is governed by $W_{1/2}$, resulting in the $1/f_{\mathrm{m}}$ dependence of Eq. (27). This dependence resembles that in BOCDA and BOCDR [13], although the relevant spectral width is different. In Brillouin sensing, it is set by the Brillouin gain bandwidth, whereas in OCDR it is determined by the source linewidth and receiver bandwidth. At high $f_{\mathrm{m}}$, adjacent spectral components are separated from the detection band, leaving only the zeroth-order component. The resulting saturation value in Eq. (11) is therefore independent of $B$ and $\delta\nu$. Its agreement with the measured high $f_{\mathrm{m}}$ limit also supports the use of electrical power, rather than beat-current amplitude, to define the spatial response in the present detection scheme.

The discrepancy at RBW = 10 MHz indicates that the effective spectral response cannot be described fully by the idealized Voigt model under the present measurement conditions. The independently measured RBW profile was close to Gaussian, as expected for a commercial spectrum analyzer [23], but the unmodulated beat spectrum was narrower than the Voigt profile calculated from the measured $B$ and $\delta\nu$. A systematic error in $\Delta f$ is unlikely to explain this difference because the high $f_{\mathrm{m}}$ saturation value, which depends on $\Delta f$ but not on $B$, $\delta\nu$, or the filter shape, agreed with the measurement within approximately 3%. The RBW FWHM was also characterized independently using a reference tone.

A residual discrepancy of approximately 10 to 20% remained at RBW = 10 MHz even when the measured filter function was used. The measured unmodulated beat spectrum was narrower than the Voigt prediction over most of the measured frequency range, indicating that the difference cannot be explained solely by the nominal RBW or by the assumed Gaussian filter shape. The origin of the remaining discrepancy could not be identified within the present study. Possible contributions include measurement bias under the actual signal conditions and optical noise associated with the present configuration [10]. Further measurements of the individual beat components will be required to clarify this discrepancy.

The present formulation also quantifies the trade-off between spatial resolution and measurement range in OCDR. Because the measurement range is inversely proportional to $f_{\mathrm{m}}$, increasing $f_{\mathrm{m}}$ improves $\Delta z$ in the tail region while reducing the accessible range. Equation (27) further shows that the number of resolvable points in this region is independent of $f_m$ and is determined by $\Delta f$ and $W_{1/2}$. Thus, increasing $\Delta f$ or reducing the effective spectral width can increase the number of resolvable points without changing this basic trade-off. The present experiments were performed with a single light source and evaluated $\Delta z$ at RBW = 1 and 10 MHz, so the predicted dependence on source linewidth and RBW has not yet been tested systematically over a wider parameter range. In particular, the effect of reducing $\delta\nu$ remains to be verified experimentally. Further measurements using different source linewidths, receiver bandwidths, and signal conditions should clarify the remaining discrepancy and establish the range over which the present model can be used for quantitative system design.

## CONCLUSION

We developed a theoretical description of the modulation frequency dependence of the spatial resolution in OCDR with a frequency shifter. By evaluating the electrical power measured by the ESA and including the source linewidth $\delta\nu$ and resolution bandwidth $B$, the model predicts two regimes: $\Delta z$ decreases approximately as $1/f_{\mathrm{m}}$ at low $f_{\mathrm{m}}$ and approaches $\Delta z_{\mathrm{sat}} = 0.563c/(\pi n \Delta f)$ at high $f_{\mathrm{m}}$. The measurements at RBW = 1 MHz reproduced this transition over a wide range of correlation orders, including the predicted saturation value. At RBW = 10 MHz, the Voigt model overestimated $\Delta z$, whereas direct use of the measured unmodulated beat spectrum reduced the discrepancy to approximately 10 to 20%. The results show that the source linewidth and receiver response must be included when predicting the spatial resolution in the low $f_{\mathrm{m}}$ region, while the high $f_{\mathrm{m}}$ limit is determined only by $\Delta f$. The model also provides a quantitative relation between spatial resolution and measurement range. Further measurements with different source linewidths and receiver bandwidths are needed to test the model over a wider range of conditions and to clarify the remaining discrepancy.

## ACKNOWLEDGMENTS

**Funding:** This work was partly supported by JSPS KAKENHI (21H04555, 24KJ0908, and 26H02136) and JST ACT-X (JPMJAX25M9) and by a research grant from the Asahipen Hikari Foundation.

**Author contributions:** Conceptualization: K.M. and T.K. Methodology: K.M. Validation: K.M., T.K., and Y.M. Investigation: K.M. Resources: Y.M. Writing—original draft preparation: K.M. and Y.M. Writing—review and editing: K.M., T.K., and Y.M. Visualization: K.M. Project administration: Y.M. Funding acquisition: T.K. and Y.M. All authors have read and agreed to the published version of the manuscript.

**Competing interests:** The authors declare that they have no competing interests.

## DATA AVAILABILITY

The data supporting the results of this study can be obtained from the corresponding author upon reasonable request.

## REFERENCES

[1] Barnoski MK, Rourke MD, Jensen SM, Melville RT. Optical time domain reflectometer. *Appl. Opt.* 1977;16(9):2375–2379.

https://doi.org/10.1364/AO.16.002375

[2] Koyamada Y, Imahama M, Kubota K, Hogari K. Fiber-optic distributed strain and temperature sensing with very high measurand resolution over long range using coherent OTDR. *J. Lightwave Technol.* 2009;27(9):1142–1146. https://doi.org/10.1109/JLT.2008.928957

[3] Fernández-Ruiz MR, Martins HF, Pastor-Graells J, Martin-Lopez S, Gonzalez-Herraez M. Phase-sensitive OTDR probe pulse shapes robust against modulation-instability fading. *Opt. Lett.* 2016;41(24):5756–5759. https://doi.org/10.1364/OL.41.005756

[4] Eickhoff W, Ulrich R. Optical frequency domain reflectometry in single-mode fiber. *Appl. Phys. Lett.* 1981;39(9):693–695. https://doi.org/10.1063/1.92872

[5] Koshikiya Y, Fan X, Ito F. Long range and cm-level spatial resolution measurement using coherent optical frequency domain reflectometry with SSB-SC modulator and narrow linewidth fiber laser. *J. Lightwave Technol.* 2008;26(18):3287–3294. https://doi.org/10.1109/JLT.2008.928916

[6] Soller BJ, Gifford DK, Wolfe MS, Froggatt ME. High resolution optical frequency domain reflectometry for characterization of components and assemblies. *Opt. Express* 2005;13(2):666–674. https://doi.org/10.1364/OPEX.13.000666

[7] Youngquist RC, Carr S, Davies DEN. Optical coherence-domain reflectometry: a new optical evaluation technique. *Opt. Lett.* 1987;12(3):158–160. https://doi.org/10.1364/OL.12.000158

[8] Hotate K, Kamatani O. Optical coherence domain reflectometry by synthesis of coherence function. *J. Lightwave Technol.* 1993;11(10):1701–1710. https://doi.org/10.1109/50.249913

[9] Hotate K, He Z. Synthesis of optical-coherence function and its applications in distributed and multiplexed optical sensing. *J. Lightwave Technol.* 2006;24(7):2541–2557. https://doi.org/10.1109/JLT.2006.874645

[10] Motoda K, Zhu G, Kiyozumi T, Ishimaru T, Takahashi H, Toge K, Mizuno Y. Noise-suppressed optical correlation-domain reflectometry using dual lasers. *Opt. Fiber Technol.* 2025;94:104294. https://doi.org/10.1016/j.yofte.2025.104294

[11] Liu C, Fan X, He Z. Low receiver-bandwidth quasi-distributed ultrasonic sensing enabled by crosstalk-suppressed optical coherence domain reflectometry. *J. Lightwave Technol.* 2025;43(24):11104–11111. https://doi.org/10.1109/JLT.2025.3618694

[12] Mizuno Y, Zou W, He Z, Hotate K. Proposal of Brillouin optical correlation-domain reflectometry (BOCDR). *Opt. Express* 2008;16(16):12148–12153. https://doi.org/10.1364/OE.16.012148

[13] Hotate K, Hasegawa T. Measurement of Brillouin gain spectrum distribution along an optical fiber using a correlation-based technique—Proposal, experiment and simulation. *IEICE Trans. Electron.* 2000;E83-C(3):405–412.

[14] Hotate K, Tanaka M. Distributed fiber Brillouin strain sensing with 1-cm spatial resolution by correlation-based continuous-wave technique. *IEEE Photon. Technol. Lett.* 2002;14(2):179–181. https://doi.org/10.1109/68.980502

[15] Hotate K. Brillouin optical correlation-domain technologies based on synthesis of

optical coherence function as fiber optic nerve systems for structural health monitoring. *Appl. Sci.* 2019;9(1):187. https://doi.org/10.3390/app9010187

[16] Zhang M, Bao X, Chai J, Zhang Y, Liu R, Liu H, Liu Y, Zhang J. Impact of Brillouin amplification on the spatial resolution of noise-correlated Brillouin optical reflectometry. *Chin. Opt. Lett.* 2017;15(8):080603. https://doi.org/10.3788/COL201715.080603

[17] Song KY, Choi JH. Measurement error induced by the power-frequency delay of the light source in optical correlation-domain distributed Brillouin sensors. *Opt. Lett.* 2018;43(20):5078–5081. https://doi.org/10.1364/OL.43.005078

[18] Kiyozumi T, Noda K, Zhu G, Nakamura K, Mizuno Y. Modified expression for spatial resolution in optical correlation-domain reflectometry. *IEEE Trans. Instrum. Meas.* 2024;73:7001711. https://doi.org/10.1109/TIM.2023.3346492

[19] Noda K, Lee H, Nakamura K, Mizuno Y. Brillouin optical correlation-domain reflectometry based on arbitrary waveform modulation: a theoretical study. *Opt. Express* 2021;29(9):13794–13805. https://doi.org/10.1364/OE.422873

[20] Noda K, Kiyozumi T, Mizuno Y, Set SY, Yamashita S. Reassessment of spatial resolution in Brillouin optical correlation-domain sensors. *J. Opt. Soc. Am. B* 2025;42(7):1556–1567. https://doi.org/10.1364/JOSAB.563168

[21] Kikuchi K, Lee H, Ohata R, Noda K, Inoue R, Mizuno Y. BOCDR achieving 6-mm spatial resolution at modulation frequencies close to Brillouin bandwidth. *J. Lightwave Technol.* 2026;44(7):2831–2840. https://doi.org/10.1109/JLT.2025.3640608

[22] Motoda K, Kiyozumi T, Set SY, Yamashita S, Mizuno Y. Experimental investigation of relationship between modulation frequency and spatial resolution in OCDR with frequency shifter. *Proc. SPIE* 2025;13639:136393P. https://doi.org/10.1117/12.3062698

[23] Rauscher C, Janssen V, Minihold R. *Fundamentals of Spectrum Analysis.* München: Rohde & Schwarz; 2001. ISBN 978-3-939837-01-5.

[24] Poppe GPM, Wijers CMJ. More efficient computation of the complex error function. *ACM Trans. Math. Softw.* 1990;16(1):38–46. https://doi.org/10.1145/77626.77629

[25] Noda K, Han G, Lee H, Mizuno Y, Nakamura K. Proposal of external modulation scheme for fiber-optic correlation-domain distributed sensing. *Appl. Phys. Express* 2019;12(2):022005. https://doi.org/10.7567/1882-0786/aaf416

[26] Motoda K, Mizuno Y. Ghost-peak-based estimation of modulation amplitude in optical correlation-domain reflectometry. *Sci. Rep.* 2026;16:14567. https://doi.org/10.1038/s41598-026-44272-3